\documentclass[twocolumn,showpacs,aps,prl,amsmath,amssymb,floatfix,%
superscriptaddress]{revtex4}
\usepackage{graphicx}
\usepackage{dcolumn}
\usepackage{bm}

\begin{document}

\title{The Interplay of Spin and Charge Channels in Zero Dimensional Systems}
\author{M.N.Kiselev}\altaffiliation[On leave from ]
{Institute for Theoretical Physics and Astrophysics, W\"urzburg
University, 97074 W\"urzburg, Germany} \affiliation{Physics
Department, Arnold Sommerfeld Center for Theoretical
Physics and Center for Nano-Science\\
Ludwig-Maximilians Universit\"at M\"unchen, 80333 M\"unchen,
Germany}
\author{Yuval Gefen}
\affiliation{Department of Condensed Matter Physics, The Weizmann
Institute of Science, Rehovot 76100, Israel }
\date{\today}
\begin{abstract}
We present a full fledged quantum mechanical treatment of  the
interplay between  the charge and the  spin zero-mode interactions
in quantum dots. Quantum fluctuations of the  spin-mode suppress
the Coulomb blockade and give rise to non-monotonic behavior near
this point.  They also greatly  enhance  the  dynamic spin
susceptibility. Transverse fluctuations become important as one
approaches the Stoner instability.  The non-perturbative effects
of zero-mode interaction are described in terms  of charge
($U(1)$)  and spin ($SU(2)$)  gauge bosons.
\end{abstract}
\pacs{73.23.Hk,73.63.Kv,75.75.+a,75.30.Gw}

\maketitle The importance of electron-electron interactions is
emphasized in low-dimensional conductors. In one-dimension
interactions in the charge and spin channels are separable
Considering  zero-dimensional quantum dots (QDs), the "Universal
Hamiltonian" \cite{Hamiltonian, review} scheme provides a
framework to study the leading interaction modes: zero-mode
interactions in the charge, spin (exchange) and Cooper channels.
While  this  Hamiltonian is simple, the physics involved is not at
all trivial. The charge channel interaction leads to the
phenomenon of the Coulomb blockade (CB). The exchange interaction
leads to  Stoner instability \cite{stoner_bulk}, which, in
mesoscopic systems as opposed to bulk, is modified
\cite{Hamiltonian}.
 Attention has been  given to the intriguing interplay
between the charge and the spin channels. This is manifest, for
example, in the suppression of certain Coulomb peaks due to
"spin-blockade" \cite{spin_blockade}. In a recent theoretical
study \cite{Alhassid} the  effect of the spin channel on Coulomb
peaks has been analyzed  employing a master equation in the
classical limit.

In this Letter we study  both transport through a metallic grain
and the dynamic magnetic susceptibility of the latter.
Specifically we find that (i) the spin modes renormalize the CB,
thus modifying the tunneling density of states (TDoS) of (hence
the differential  conductance through) the dot (cf. Fig.2 and Eq.
\ref{densitos}).  For an Ising-like spin anisotropy (represented
by $1-\epsilon$) the longitudinal mode partially suppresses  the
CB. Transverse modes act qualitatively in the same way, but as one
approaches the Stoner instability point (from the disordered
phase),  the effect of transverse fluctuations reverses its sign
and acts towards suppressing the conductance (i.e., {\it
enhancing} the CB). This results in a non-monotonic behavior of
the TDoS.  (ii) The longitudinal magnetic susceptibility
$\chi^{zz}$ (\ref{chi2})  diverges at the thermodynamic Stoner
Instability point, while $\chi^{+-}$   (\ref{chi2}) is enhanced
but  remains finite. However, one notes that the static transverse
susceptibility is enhanced by the gauge fluctuations.

Our study is the first full fledged quantum mechanical analysis of
spin fluctuations and the charge-spin interplay in zero
dimensions. The non-perturbative effects of zero-mode charge
interaction (e.g. zero-bias anomaly \cite{zba}) are described in
terms of the propagation of gauge bosons ($U(1)$ gauge field)
\cite{Kam96}. Here we adopt similar ideas to account for spin
fluctuations described by the non-abelian $SU(2)$ group. The
Coulomb and longitudinal spin components are accounted for
"exactly", while transverse spin fluctuations are analyzed
perturbatively (with the easy-axis anisotropy, $\epsilon$
\cite{anisotropy}). These fluctuations become important as one
approaches the Stoner instability.  Here we restrict ourselves to
the Coulomb valley regime and ferromagnetic exchange interaction.

Before proceeding we recall that beyond the  thermodynamic Stoner
instability point, $J_{th}=\Delta$ ($\Delta$ being the mean level
spacing)  the spontaneous magnetization is an extensive quantity.
At smaller values of the exchange coupling,
 $J_{c}$$<$$J<J_{th}$, finite magnetization
shows up \cite{Hamiltonian}, which, for finite systems,  does not
scale linearly with the size of the latter \cite{com3}. Its
non-self-averaging nature gives rise to strong sample-specific
mesoscopic fluctuations. The incipient instability for finite
systems  is given by  $J_{c}=\Delta/(1+\epsilon)$ for an even
number of spins in the dot and $J_{c}=\Delta/(1+\epsilon/2)$ for
an odd number  \cite{com2}.

{\it Hamiltonian and correlators}. Our QD is taken to be in the
metallic regime, with its internal dimensionless conductance $g\gg
1$. Discarding both Cooper and spin-orbit interaction channels,
the description of our metallic QD allows for  only two other
channels, namely charge and spin. While the charge interaction is
invariant under $U(1)$ transformation, the spin interaction
possesses a non-abelian $SU(2)$ symmetry associated with the
non-commutativity of the quantum spin components.

The Universal Hamiltonian now assumes the form
\begin{equation}
H=\sum_{\alpha,\sigma}\epsilon_\alpha
a^\dagger_{\alpha,\sigma}a_{\alpha,\sigma} +H_C+H_S \label{Ham1}
\end{equation}
Here $\alpha$ denotes a single-particle  orbital state  with spin
projection $\sigma$. For simplicity, below  we confine ourselves
to the GUE case. The Hamiltonian
$H_C=E_c\displaystyle\left(n-N_0\right)^2$ accounts for the
Coulomb blockade, $E_c=e^2/2C$ is a charging energy, $n$ the
number operator; $N_0$ stands for a positive background charge
tuned to a Coulomb valley. The Hamiltonian
\begin{equation}
H_{S}=-J\left[\left(\sum_\alpha S_\alpha^z\right)^2
+\epsilon\left\{\left(\sum_\alpha
S_\alpha^x\right)^2+\left(\sum_\alpha
S_\alpha^y\right)^2\right\}\right]
\end{equation}
represents the spins
$\vec{S}_{\sigma\sigma^{'}}$$\equiv$$\frac{1}{2} \sum_\alpha
a^\dagger_{\alpha,\sigma} \vec{\sigma}_{\sigma\sigma^{'}}
a_{\alpha,\sigma^{'}}$ interaction within the dot. Hereafter we
assume strong  easy axis anisotropy \cite{anisotropy},
$\epsilon=J_\perp/J_\parallel < 1$. In this case the spin rotation
symmetry is reduced to $SO(2)$. We will treat the terms of
transverse and longitudinal (Ising) fluctuations independently.

The Euclidian action for the model (\ref{Ham1}) is given by
\begin{equation}
S=\int_0^\beta {\cal L}(\tau) d\tau=
\int_0^\beta\left[\sum_{\alpha\sigma}\bar\psi_{\alpha\sigma}(\tau)[
\partial_\tau
+\mu]\psi_{\alpha\sigma}(\tau) -H\right] d\tau \label{act1}
\end{equation}
Here $\psi$ stand for Grassmann variables describing electrons in
the dot. The imaginary time single particle Green's function (GF)
is written as
\begin{equation}
{\cal G}_{\alpha\sigma\sigma'}(\tau_i,\tau_f)=\frac{1}{Z(\mu)}\int
D[\bar\psi\psi]
\bar\psi_{\alpha\sigma}(\tau_i)\psi_{\alpha\sigma'}(\tau_f)e^{S[\bar\psi\psi]}
\label{GF1}
\end{equation}
where partition function $Z(\mu)$ is given by
\begin{equation}
Z(\mu)= \int D[\bar\psi\psi] e^{S[\bar\psi\psi]}. \label{Parti}
\end{equation}

 Employing a Hubbard-Stratonovich transformation with the
bosonic fields $\phi$ (for charge) and $\vec\Phi$ (for spin)
\begin{widetext}
\begin{eqnarray}
\exp\left(-\int_0^\beta d\tau H_C\right)&=&\int
D[\phi]\exp\left(-\int_0^\beta d\tau\left[
\frac{\phi^2}{2V}-i\phi\left[\sum_\alpha
n_\alpha - N_0\right]\right]\right) \nonumber\\
\exp\left(-\int_0^\beta d\tau H_S\right)&=&\int
D[\vec{\Phi}]\exp\left(-\int_0^\beta d
\tau\left[\frac{\vec{\Phi}^2}{J}
-\Phi^z\sum_\alpha(n_{\alpha\uparrow}-n_{\alpha\downarrow})
-\sqrt{\epsilon}\left(\Phi^+\sigma^-+\Phi^-\sigma^+\right)\right]\right)
\end{eqnarray}
\end{widetext}
we obtain a Lagrangian which includes a term quadratic in $\Psi$
 ${\cal L}_\Psi=\sum_\alpha
\bar\Psi_\alpha M_\alpha\Psi_\alpha$. Here we have  used a spinor
notation $\bar \Psi_\alpha=(\bar\psi_{\uparrow\alpha}
\bar\psi_{\downarrow\alpha})$ and the matrix $M_\alpha$ is given
by
\begin{eqnarray}
M_\alpha= \left(
\begin{array}{cc}
\partial_\tau -\xi_\alpha +i\phi+ \Phi^z & \sqrt{\epsilon}\Phi^-\\
\sqrt{\epsilon}\Phi^+ &\partial_\tau -\xi_\alpha+i\phi - \Phi^z
\end{array}
\right).
\label{U1b}
\end{eqnarray}
Our goal here is to obtain the GF. We first add source fields to
the Lagrangian
\begin{equation}
{\cal L}_{\Lambda \Upsilon}={\cal L}+ \bar \Lambda \Psi +\bar \Psi
\Lambda+\vec{\Upsilon}\vec{\Phi},\nonumber
\end{equation}
and define the generating function ${\cal Z}$ as follows
\begin{equation}
{\cal Z}= Z(\mu)^{-1}\int D[\bar \Psi \Psi] D[\vec{\Phi}\phi]
\exp\left(\int_0^\beta d\tau {\cal
L}_{\Lambda\Upsilon}(\tau)\right),
\end{equation}
where $Z(\mu)$ is a partition function (\ref{Parti}) of the dot.
The fermionic ($2$$\times$$2$) and bosonic ($3$$\times$$3$) matrix
GFs are:
\begin{equation}
{\cal
G}^{\sigma\sigma'}_\alpha(\tau_i,\tau_f)=\frac{\partial^2{\cal Z}}
{\partial \bar\Lambda^\sigma_{\tau_{f}}\partial
\Lambda^{\sigma'}_{\tau_{i}}},\; {\cal
D}^{\mu\nu}(\tau_i,\tau_f)=\frac{\partial^2{\cal Z}} {\partial
\Upsilon^\mu_{\tau_{f}}\partial
\Upsilon^\nu_{\tau_{i}}}\label{grefunc}
\end{equation}
with $\Lambda \to 0, \vec{\Upsilon} \to 0$. Here ${\cal G}_\alpha$
is given by   ${\cal G}_{\alpha\sigma}=-\langle T_\tau
\Psi_{\alpha\sigma}(\tau_f)\bar\Psi_{\alpha\sigma}(\tau_i)\rangle$
while ${\cal D}^{\mu\nu}=-\langle T_\tau
\Phi^\mu(\tau_i)\Phi^{\nu}(\tau_f)\rangle$.

{\it Gauge transformation}. We now apply a (non-unitary)
transformation to gauge out both the Coulomb and the longitudinal
part of the spin interaction $\tilde M_\alpha = W M_\alpha
W^{-1}$. We have $\tilde \Psi$$=$$ W(\tau)$$\Psi$ and
$\tilde{\bar{\Psi}}$$=$$\bar \Psi$$ W^{-1}(\tau)$ with
\begin{eqnarray}
W(\tau)= e^{i\theta(\tau)}\left(
\begin{array}{cc}
e^{\eta(\tau)} &0 \\
0&e^{-\eta(\tau)}
\end{array}
\right). \label{U1c}
\end{eqnarray}
Here $\theta$ ($\eta$) accounts for the $U(1)$ fluctuations of the
charge (longitudinal) fluctuations,
\begin{equation}
\theta=\int_0^\tau\left(\phi(\tau')-\phi_0\right)d\tau',\;
\eta=\int_0^\tau\left(\Phi^z(\tau')-\Phi^z_0\right)d\tau'.
\label{te}
\end{equation}
In defining the gauge fields $\phi_0$  \cite{Kam96} and $\Phi^z_0$
one needs to account  for possible  winding numbers $(k,m$$=$$0$$\pm1$$,.
..$$)$ \cite{Efetov}:
\begin{equation}
\beta\phi_0=\int_0^\beta\phi(\tau)d\tau+2\pi k,\;\;
\beta\Phi^z_0=\int_0^\beta\Phi^z(\tau) d\tau + 2i\pi m
\label{windings}
\end{equation}

In Eq.(\ref{te}) initial conditions $W(0)=1$ and periodic boundary
conditions   $W(0)=W(\beta)$ are used. As a result, the diagonal
part of the gauged inverse electron's GF ($\tilde M_\alpha$) does
not depend on the finite frequency components of fields. The
off-diagonal part can be taken into account by a perturbative
expansion in $\epsilon < 1$. We represent $\tilde M_\alpha =
({\cal G}_\alpha^{[0]})^{-1}+\Sigma_\Phi$ with
$\left({\cal G}^{[0]}_{\alpha}(\tau)\right)^{-1}=(\partial_\tau
-\xi_\alpha +i\phi_0)\hat 1 +\Phi_0^z \sigma^z$
and the self-energy
$\Sigma_\Phi=\sqrt{\epsilon}\left(\Phi^-e^{2\eta}\sigma^+
+\Phi^+e^{-2\eta}\sigma^-\right)$. We next calculate the Green's
function
\begin{eqnarray}
{\cal
G}^{\sigma\sigma'}_{\alpha}(\tau,\tilde\mu)=\delta_{\sigma\sigma'}\langle{\
\cal G}^{[0]}_\alpha(\tau,\tilde\mu) \exp\left(\sigma
\eta_{\tau}+i\theta_{\tau}\right)\rangle_{\vec{\Phi}\phi}.
\label{GF2}
\end{eqnarray}
Hereafter $\langle...\rangle_{\vec{\Phi}\phi}$
($\langle...\rangle_{tr}$) denotes Gaussian averaging over
fluctuations (transverse fluctuations) of the bosonic field
$(\vec{\Phi},\phi)$ and
$\tilde\mu$$=$$\mu$$+$$\sigma\Phi^z_0$$+$$i\phi _0$. Integrating
over all Grassmann variables and expanding $\tilde M_\alpha$ with
respect to the transverse fluctuations, one obtains \cite{com4}
\begin{widetext}
\begin{eqnarray}
\nonumber &&{\cal
G}_{\alpha\sigma}(\tau_i,\tau_f)=\frac{1}{Z(\mu)}\int
d\vec{\Phi}_0 d\phi_0 \exp\left(-\frac{(\Phi^z_0)^2}{T
J}-\frac{(\phi_0)^2}{4TE_{c}} -\beta\Omega_0 \right)
\int\prod_{n\ne 0} d\Phi^z_n d\phi_n\exp\left\{T\sum_{n\ne
0}\left(-\frac{\Phi^z_n\Phi^z_{-n}}{J}
-\frac{\phi_n\phi_{-n}}{4E_{c}}\right)\right\}\nonumber\\
&&W(\tau_i)\left[{\cal
G}^{[0]}_\alpha(\tau_i-\tau_f,\mu+i\phi_0+\sigma\Phi^z_0)-
\sum_{k=1}^{\infty}\int d\tau_1 ...\int d\tau_{2k}\langle({\cal
G}^{[0]}_{\alpha} \Sigma)^{2k}\rangle_{tr} {\cal
G}^{[0]}_\alpha(\tau_{2k}-\tau_f,\mu+i\phi_0+\sigma\Phi^z_0)\right]
W^{-1}(\tau_f).
\nonumber\\
\label{grf}
\end{eqnarray}
\end{widetext}
where $\Omega_0(\tilde\mu)= -T\ln Z_0$, $Z_0$ is the partition
function of the non-interacting electron gas. Also, computing the
bosonic correlator (Eq (\ref{grefunc})), we find
\begin{eqnarray}
{\cal D}^{\mu\nu}(\tau_i,\tau_f)=\frac{1}{Z(\mu)}\int
D[\vec{\Phi}] \;\;\Phi^\mu(\tau_i)\Phi^\nu(\tau_f)\times\nonumber\\
\times\exp\left({\tt Tr}\log\left[1+{\cal
G}^{[0]}_\alpha\Sigma_\Phi\right]-\frac{1}{J}\int_0^\beta
\vec{\Phi}^2 d\tau\right). \label{GF2}
\end{eqnarray}
\begin{figure}
\includegraphics[width=0.2\textwidth]{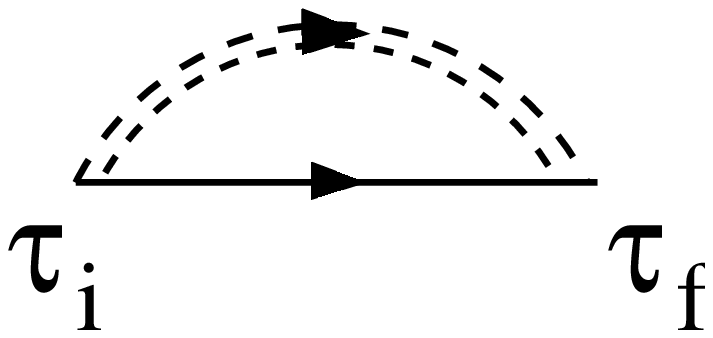}
\includegraphics[width=0.3\textwidth]{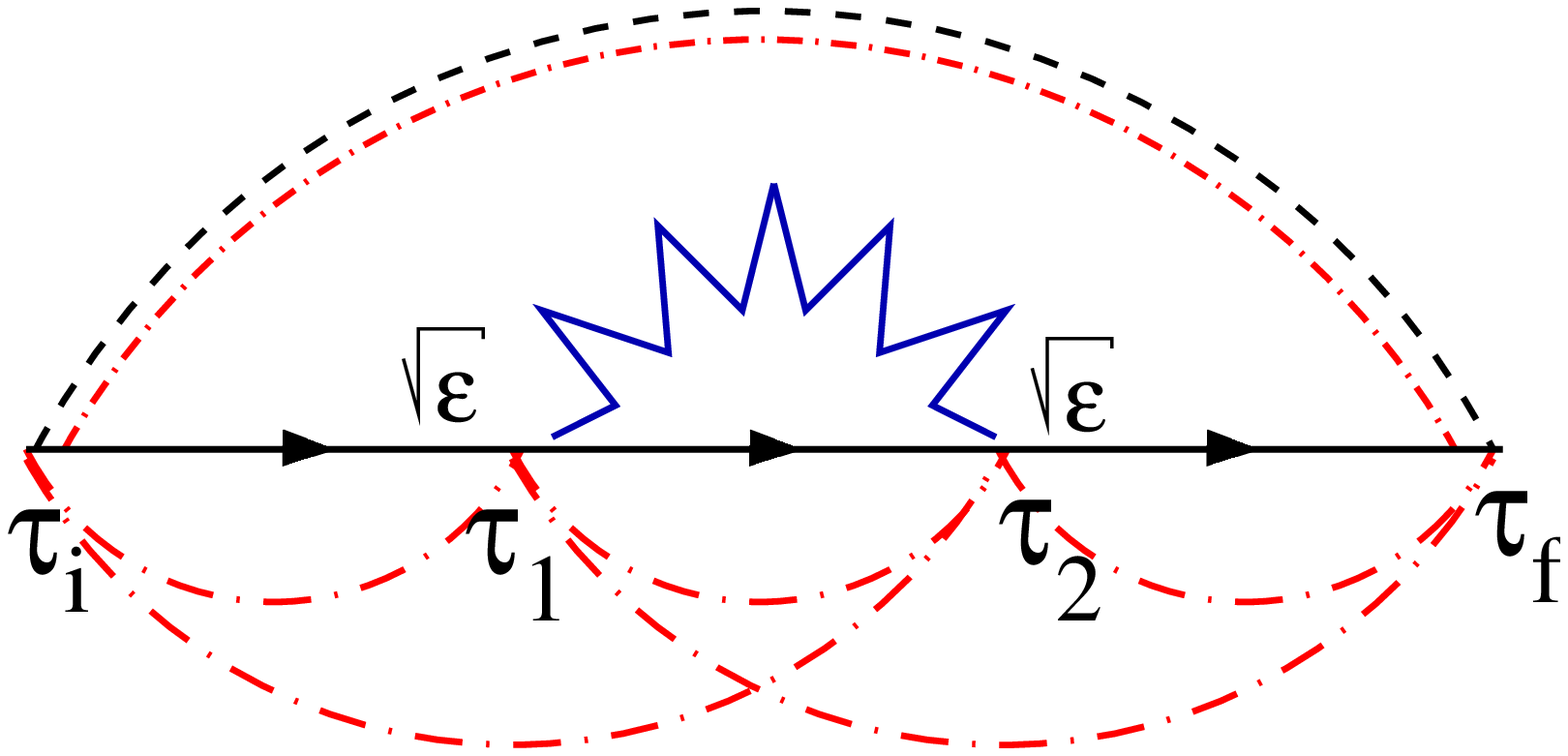}\\
\caption{\label{fig:f2} First and second order Feynman diagrams
contributing to electron's GF. Solid line represents ${\cal
G}^{[0]}_{\alpha,\sigma}$; double  dashed line stands for
combination of Coulomb and longitudinal bosons; single dashed line
denotes a longitudinal boson while the zig-zag line represents
$\langle\Phi^+(\tau_1)\Phi^-(\tau_2)\rangle$.}
\end{figure}
In the spirit of \cite{Kam96}, the interaction of electrons with
the finite-frequency charge and longitudinal modes ($\phi_n$,
$\Phi^z_n$)  may be interpreted in terms of a gauge boson
\cite{com5} dressing the electron propagator (cf Fig.1a). The {\it
exact} electronic GF (depending on winding numbers \cite{Efetov})
is given by \cite{Kam96}
\begin{equation}
{\cal G}_{\alpha,\sigma}(\tau_i-\tau_f)={\cal
G}^{[0]}_{\alpha,\sigma}(\tau_i-\tau_f,
\tilde\mu)e^{-S_\parallel(\tau_i-\tau_f)}, \label{GF4}
\end{equation}
where the Coulomb-longitudinal $U$$($$1$$)$ gauge factor is
$$
S_\parallel(\tau)=4T\sum_{n\ne
0}\frac{E_c-J/4}{\omega_n^2}\sin^2\left(\frac{\omega_n\tau}{2}\right)=
$$
\vspace*{-4mm}
\begin{equation}
=\left(E_c-\frac{J}{4}\right)\left(|\tau|-\frac{\tau^2}{\beta}\right).
\label{GF5}
\end{equation}
The exchange interaction effectively modifies the charging energy.
For long-range interaction this correction is small
$E_c/J$$\sim$$(k_F L)^{d-1}$ \cite{review} ($L$ is a linear size
of the $d$-dimensional confined electron gas, $k_F$ is a Fermi
momentum), while for contact interaction $E_c-J/4=0$
\cite{review}.

{\it Transverse fluctuations}.
 The first non-vanishing diagram of our  expansion (\ref{grf}) is depicted in
Fig.1b. Here
\begin{equation}
{\cal
G}^{[0]}_{\alpha,\sigma}(\tau)=e^{-\xi_{\alpha\sigma}\tau}\left(n_{
\xi_{\alpha\sigma}}(1-\theta_\tau)-
(1-n_{\xi_{\alpha\sigma}})\theta_\tau\right) \;\;\;\; \label{ferm}
\end{equation}
the transverse correlator is considered in the Gaussian
approximation
\begin{equation}
\langle\Phi^+(\tau_1)\Phi^-(\tau_2)\rangle
=\frac{J}{2}\delta\left(\tau_1-\tau_2\right) +\frac{\epsilon
J^2}{\displaystyle 2\beta(\Delta-\epsilon J)}. \label{dd}
\end{equation}
In   Eq.(\ref{dd}) the first term is a manifestation of the white
noise fluctuations of the fields $\vec{\Phi}$ arising from the
Gaussian weight factor (cf. Eq. (\ref{GF2})). The second term is
related to the non-Gaussian factor in  ${\cal D}^{\mu\nu}$ and
reflects the feed-back of (the  ${\cal D}$-dependent) ${\cal
G}^{[0]}$ on ${\cal D}$. Note that the transverse components
$\Phi^{\pm}$ are always accompanied by the gauge factors $e^{\mp
2\eta}$, hence the longitudinal bosons contribute
to the dynamics involving the transverse fluctuations.\\

To proceed we now sum Eq.(\ref{grf})  over $\alpha$. Perturbative
corrections  to the electron GF coming from transverse
fluctuations are now expanded in $\epsilon$ and summed up in the
factor $F_\perp$
\begin{equation}
G(\tau)=G_0(\tau)e^{-S_{\parallel}(\tau)}F_\perp(\tau,\epsilon),
\label{GFandgauge}
\end{equation}
where
\begin{equation}
F_\perp(\tau,\epsilon)=1+\sum_{n,m}\{f^{(n)}g^{(m)}\}.
\label{expansion}
\end{equation}
The effects of disorder are incorporated in the bare density of
states $\nu_0=1/\Delta$. We denote
$\tau$$\equiv$$\tau_f$$-$$\tau_i$. At finite temperatures we
employ the conformal transformation
$1/(\tau_i-\tau_f)\to\pi/(\beta\sin\left(\pi(\tau_i-\tau_f)/\beta\right))$.
The factors $\{f^{(n)}g^{(m)}\}$ refer to  diagrams containing
$n+m$ ${\cal D}$ (zig-zag) line correlators,  in $n$ $(m)$ of
which we employ  the first, $\delta$-function (second, constant)
term of Eq. (\ref{dd}). The $\{f^{(n)}g^{(m)}\}$ factor is $\sim
\epsilon^{n+2m}$. Here we calculate   $F_\perp$ to order $
\epsilon^2$.

In general,  we may write the effective transverse gauge boson as
$D_\perp(\epsilon,\tau)=\exp\left(-S_\perp(\epsilon,\tau)\right)$
\cite{exponentiation}. $S_\perp$ preserves the symmetry (in
$\tau$) with respect to $\beta/2$ to all orders of $\epsilon$. It
can therefore be written as $S_\perp(\epsilon,
\tau-\frac{\tau^2}{\beta})$.

The first term in (\ref{expansion}) (of order $\epsilon$) is given
by
\begin{equation}
f^{(1)}=\epsilon \frac{J}{2}\left(\tau-\frac{\tau^2}{\beta}\right)
\label{f1sing}
\end{equation}
This contribution is of the same origin as that of the
longitudinal boson  part  (namely it comes from the
$\frac{1}{J}\int_0^\beta \vec{\Phi}^2 d\tau $ term in Eq.
(\ref{dd}).)  Along with the other $m=0$ terms in
(\ref{expansion}) it can be exponentiated \cite{exponentiation},
resulting in $J/4\to J(1+2\epsilon)/4$ in the expression for
$S_\parallel$. For the isotropic model one obtains
$(1+2\epsilon)/4 \to S(S+1)$.

 There are contributions to
$S_\perp$ arising from the second term of  Eq.(\ref{dd}).
 As a result, the  lowest, $\sim \epsilon^2$ contribution
(the $\{f^{(0)}g^{(1)}\}$ term of the expansion (\ref{expansion}))
leads to a non-Gaussian  contribution to $ S_\perp$ which is $\sim
- \frac{\epsilon^2 J^2}{\displaystyle 2\beta(\Delta-\epsilon J)}
[\tau -\frac{\tau^2}{\beta}]^2 $.  It is easy to show that below
the incipient Stoner instability, $J$$<$$J_{c}$, $S_\perp$ is
dominated by the "white noise" term of Eq.(\ref{dd}), while above
this point it is the second (singular Stoner) term in (\ref{dd})
which dominates.

{\it Tunneling density of states}.  The conductance $g_T$ is
related to the tunnelling DoS $\nu$ through $ g_T=\displaystyle
\frac{e}{\hbar}\int d\epsilon
\nu(\epsilon)\Gamma(\epsilon)\left(-\frac{\partial f_F}{\partial
\epsilon}\right) $ where $f_F$ is the Fermi distribution function
at the  contact and $\Gamma$ is the golden rule dot-lead
broadening.  To obtain the TDoS from the GF,
Eq.(\ref{GFandgauge}), we deform the contour of integration in
accordance with \cite{Kam96}. As a result, the  TDoS is given by
\cite{Mat02}
\begin{equation}
\nu(\epsilon)=-\frac{1}{\pi}\cosh\left(\frac{\epsilon}{2T}\right)
\int_{-\infty}^\infty \sum_\sigma\langle
G_\sigma\left(\frac{1}{2T}+it\right)\rangle_{k,m}e^{i\epsilon t}
dt \label{densitos}.
\end{equation}
where $\langle...\rangle_{k,m}$ denotes a summation over all
winding numbers for Coulomb and longitudinal zero-modes
\cite{Efetov}. We have computed the temperature and energy
dependence of the TDoS for various values of $\epsilon$. These are
depicted in Fig.2. The energy dependent TDoS shows an intriguing
non-monotonic behavior at energies comparable to the charging
energy $E_c$. This behavior, absent for $J=0$, is due to the
contribution of the second term in Eq.(\ref{dd}). It is amplified
in the vicinity of the Stoner point, and signals the effect of
collective spin excitations (incipient ordered phase).

{\it Spin susceptibilities}. The spin susceptibilities are defined
through
\begin{equation}
\chi^{\mu\nu}(\tau_i,\tau_f)= \frac{\partial^4{\cal Z}} {\partial
\bar\Lambda^\mu_{\tau_{f}}\partial \Lambda^\nu_{\tau_{f}}\partial
\bar\Lambda^\nu_{\tau_{i}}\partial \Lambda^\mu_{\tau_{i}}}
\label{chi1}
\end{equation}
The longitudinal susceptibility ($\chi^{zz})$ is {\it not}
affected by the gauge bosons.  By contrast,
 the transverse
$\chi^{+-}$ acquires the gauge factor $\langle$$
e^{2\eta(\tau)}$$\rangle_{m,\Phi^z}$, where the average is
performed with respect to the Gaussian fluctuations of $\Phi^z$
and, in principle, the  winding numbers (cf. Eq.(\ref{windings})).
In practice, since $T$$>$$J$, only the $m=0$ winding should be
taken into account; $T$$>$$\Delta$ allows us to evaluate the path
integral in the Gaussian approximation. One finds to  leading
order in $\epsilon$
\begin{equation}
\chi^{zz}(\tau)=\frac{\chi_0}{1-J\chi_0},\;\;\;\;\;
\chi^{+-}(\tau)=\frac{\epsilon \chi_0 e^{J\tau}}{1-\epsilon
J\chi_0} \label{chi2}
\end{equation}
where $\chi_0=1/\Delta$. The above susceptibilities are given as
function of $\tau$. To obtain the dynamic susceptibilities one
needs to Fourier transform and then continue to real frequencies.
$\chi^{zz}$ (\ref{chi2}) \cite{shek} diverges at the thermodynamic
Stoner Instability point, while $\chi^{+-}$   remains finite.
Notwithstanding,  the static transverse susceptibility is enhanced
by the gauge fluctuations. The dynamic behavior (including
relaxation processes)   and the $\epsilon$ corrections to
$\chi^{\mu\nu}$ will be discussed elsewhere.\cite{unp}.

\begin{figure}
\includegraphics[width=0.4\textwidth]{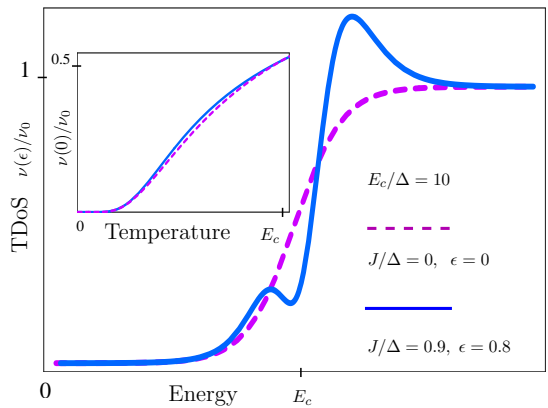}
\caption{\label{fig:f3} The spin-normalized  tunneling density of
states shown as function of energy. Insert: TDoS as function of
temperature. }
\end{figure}
Summarizing, we investigate influence of spin and charge zero-mode
interactions on the TDoS and the  susceptibilities. Longitudinal
spin fluctuations suppress the CB and the static longitudinal
susceptibility is greatly enhanced near the Stoner instability.
Transverse fluctuations generally tend to suppress the CB, but
also contain a term which dominates the dynamics  near the Stoner
instability and {\it enhances} the CB. The transverse
susceptibility will be enhanced as well.  On a more technical
level,  Coulomb interaction is described in terms of Abelian
(U(1)) gauge theory and lead to Gaussian gauge factor, the spin
interaction, being a subject of non-Abelian (SU(2)) gauge gives
rise to non-Gaussian gauge factors.

We are acknowledge useful discussions with Y.Alhassid, L.Glazman,
I.V. Lerner,  K.Matveev, A.Mirlin and Z.Schuss. We acknowledge
support  by SFB-410 grant, the Transnational Access program
RITA-CT-2003-506095 (MK), an ISF grant of the Israel Academy of
Science, the EC HPRN-CT-2002-00302-RTN and the AvH Foundation
(YG).  We are grateful to Argonne National Laboratory for the
hospitality during our visit. Research in Argonne was supported by
U.S. DOE, Office of Science, under Contract No. W-31-109-ENG-39.


\begin{thebibliography}{99}
\bibitem{com3}
As both $J$ and $\Delta$ are inversely proportional to the volume
of the system, the spin of the ground state  scales as
$S_g\sim\left[\epsilon J/(\Delta- J)\right]$.
\bibitem{com2} This criterion generalizes the results of Ref. \cite{Hamiltonian} for the case
of an easy-axis  anisotropy. In the isotropic case it reads
$J^{even}_{c}=\Delta/2$, $J_{c}^{odd}=2\Delta/3$, whereas for the
Ising case one has $J^{even}_{c}=J_{c}^{odd}=J_{th}=\Delta$.
\bibitem{com5} The exponential gauge factor being continuous at points $\tau=0,\beta$ does
not, strictly speaking, correspond to a bosonic propagator. We,
nevertheless, keep the terminology of \cite{Kam96}.
\bibitem{exponentiation} The transverse gauge boson is given by
$D_\perp(\tau)$$=$$\exp$$\left(-S_\perp(\tau)\right)$, where
 $S_\perp(\tau)$$=$$-\ln$$\left[1+\sum_{n,m}\{f^{(n)}g^{(m)}\}\right]$.
\end{thebibliography}
\end{document}